\documentclass[11pt,twoside]{article}
\usepackage{amsmath}
\usepackage{amssymb}
\usepackage{amscd}
\usepackage{mathrsfs}
\usepackage{upgreek}
\usepackage[fleqn,tbtags]{mathtools}
\DeclareMathAlphabet{\eurm}{U}{eur}{m}{n}
\DeclareMathAlphabet{\eubf}{U}{eur}{b}{n}
\usepackage{pict2e}
\usepackage{graphicx}
\usepackage{epstopdf}
\usepackage{boxedminipage}
\newcommand{\comment}[1]{}
\newenvironment{mybulist}{\begin{list}{$\bullet$}%
{\setlength{\leftmargin}{10pt}
  \setlength{\topsep}{6pt}
  \setlength{\parsep}{-2pt}}}{\end{list}}
\newcommand{\onto}{\rightarrowtail}
\newcommand{\into}{\hookrightarrow}
\newcommand{\ten}[1]{\operatorname*{\otimes}_{\!{\scriptscriptstyle #1}} }
\newcommand{\dir}[1]{\operatorname*{\oplus}_{\!{\scriptscriptstyle #1}} }
\newcommand{\cart}[1]{\operatorname*{\times}_{\!{\scriptscriptstyle #1}} }
\newcommand{\medoplus}{\operatornamewithlimits{\textstyle{\bigoplus}}}
\newcommand{\medotimes}{\operatornamewithlimits{\textstyle{\bigotimes}}}
\newcommand{\Aut}{\operatorname{Aut}}
\newcommand{\End}{\operatorname{End}}
\newcommand{\CO}{\mathrm{C}}
\newcommand{\LO}{\mathrm{L}}
\newcommand{\TO}{\mathrm{T}}
\newcommand{\VO}{\mathrm{V}}
\newcommand{\TS}{\TO^{*}\!}
\newcommand{\JO}{\mathrm{J}}
\newcommand{\TM}{\TO{\sst\!}\M}
\newcommand{\JE}{{\JO\E}}
\newcommand{\TE}{\TO{\sst\!}\E}
\newcommand{\VE}{{\VO{\sst\!}\E}}
\newcommand{\RR}{{\mathbb{R}}}
\newcommand{\CC}{{\mathbb{C}}}
\newcommand{\NN}{{\mathbb{N}}}
\newcommand{\LL}{{\mathbb{L}}}
\newcommand{\UU}{{\mathbb{U}}}
\newcommand{\VV}{{\mathbb{V}}}
\newcommand{\ZZ}{{\mathbb{Z}}}
\newcommand{\E}{E} 
\newcommand{\Gb}{G} 
\renewcommand{\H}{H} 
\newcommand{\K}{K} 
\newcommand{\M}{M} 
\renewcommand{\P}{P}
\newcommand{\Pm}{\P_{\!\!m}}
\let\Sec=\S
\renewcommand{\S}{S} 
\newcommand{\V}{V} 
\newcommand{\Lie}{{\mathfrak{L}}} 
\newcommand{\kg}{\upkappa}
\newcommand{\0}{{\sst0}}

\newcommand{\txtm}[1]{\mbox{$\smash{#1}$}}
\newcommand{\st}{\scriptstyle}
\newcommand{\sst}{\scriptscriptstyle}
\newcommand{\id}{{1\!\!1}}
\newcommand{\Id}[1]{{\id}\!{}_{#1}{}}
\newcommand{\weu}[1]{{\wedge^{\!#1}}}
\newcommand{\fnb}[1]{[\![#1]\!]}
\newcommand{\suc}[1]{{\{\![#1]\!\}}}    
\newcommand{\la}{\lambda}
\newcommand{\tn}{{\,\otimes\,}}
\newcommand{\we}{{\,\wedge\,}}
\newcommand{\wee}{\,{\scriptstyle\lozenge}\,}
\newcommand{\dO}{\mathrm{d}}
\newcommand{\dH}{\mathrm{d}_{\sst{\mathrm H}}}
\newcommand{\de}{\partial}
\newcommand{\zero}{{\scriptscriptstyle{(0)}}}
\newcommand{\one}{{\scriptscriptstyle{(1)}}}
\newcommand{\two}{{\scriptscriptstyle{(2)}}}
\newcommand{\Lagr}{\Lambda}
\newcommand{\comp}{\mathbin{\raisebox{1pt}{$\scriptstyle\circ$}}}
\newcommand{\lfr}{\mathfrak{l}}
\newcommand{\cj}[1]{\overline{#1}}
\newcommand{\lin}{{\scriptscriptstyle\bigstar}}
\newcommand{\alin}{{\overline{\scriptscriptstyle\bigstar}}}
\newcommand{\Sc}{\cj{\S}}

\newcommand{\T}{T}
\newcommand{\U}{U}
\newcommand{\Uc}{\cj{\U}}
\newcommand{\Ua}{\cj{\U}{}^\lin}
\newcommand{\Ul}{\U{}^{\lin}}
\newcommand{\Vc}{\cj{\V}}
\newcommand{\Val}{\V{}^\alin}
\newcommand{\Va}{\cj{\V}{}^\lin}
\newcommand{\Vl}{\V{}^\lin}
\newcommand{\W}{W}
\newcommand{\Wc}{\cj{\W}}

\newcommand{\Wl}{\W{}^\lin}
\newcommand{\X}{X}
\newcommand{\Y}{Y}

\newcommand{\YL}{\Y\!\!\spec{L}}
\newcommand{\YR}{\Y\!\!\spec{R}}
\newcommand{\Z}{Z}

\newcommand{\Zl}{\Z{}^\lin}
\newcommand{\I}{I}
\newcommand{\Ic}{\cj{\I}}
\newcommand{\Il}{\I{}^\lin}
\newcommand{\HO}{\mathrm{H}}
\newcommand{\iO}{\mathrm{i}}
\newcommand{\geu}{\eurm{g}}
\newcommand{\heu}{\eurm{h}}
\newcommand{\kD}{\eurm{k}}
\newcommand{\Reu}{\eurm{R}}
\newcommand{\Teu}{\eurm{T}}
\newcommand{\dga}{\upgamma}
\newcommand{\bang}[1]{{\langle#1\rangle}}
\newcommand{\cs}{{\mathclap{{\,\,\raisebox{2pt}{$\sst{-}$}}}\Gamma}}
\newcommand{\lu}{{\,l\hspace{-5.1pt}l\,}}
\newcommand{\cev}[1]{\smash{\overset{\smash{{}_{\gets}}}{#1}}}
\newcommand{\pint}{{\scriptscriptstyle\mathord{\rfloor}}}
\newcommand{\nasl}{{\rlap{\raise1pt\hbox{\,/}}\nabla}}
\newcommand{\speco}[1]{{\scriptscriptstyle{\textnormal{#1}}}}
\newcommand{\spec}[1]{{}_{\speco{#1}}}

\newcommand{\grav}{\spec{grav}}
\newcommand{\emag}{\spec{em}}
\newcommand{\Dir}{\spec{Dir}}

\newcommand{\Hivac}{{\scriptstyle{\mathcal{H}}}_{\sst0}}
\newcommand{\grade}[1]{{\lfloor#1\rceil}}
\newcommand{\gradezero}{{\!\sst\grade{\!0\!}}}
\newcommand{\gradeone}{{\!\sst\grade{\!1\!}}}
\newcommand{\gh}{\omega}
\newcommand{\agh}{\varpi}
\newcommand{\CCal}{\mathcal{C}}
\newcommand{\DC}{\mathcal{D}}
\newcommand{\DCo}{\DC_{\!\circ}}
\newcommand{\EC}{\mathcal{E}}
\newcommand{\FC}{\mathcal{F}}
\newcommand{\HC}{\mathcal{H}}
\newcommand{\OC}{\mathcal{O}}
\newcommand{\uOC}{\underline{\OC}}

\newcommand{\VC}{\mathcal{V}}
\newcommand{\uVC}{\underline{\VC}}
\newcommand{\VCo}{{\VC\!_{\circ}}}

\newcommand{\XC}{\mathcal{X}}
\newcommand{\YC}{\mathcal{Y}}
\newcommand{\ZC}{\mathcal{Z}}
\newcommand{\uZC}{\underline{\ZC}}
\newcommand{\abs}[1]{{\mathrm{#1}}} 
\newcommand{\tra}[1]{\abs{#1}^{\!*}}

\newcommand{\brstS}{{\scriptstyle{\mathrm S}}}
\newcommand{\sbot}{{\scriptscriptstyle\bot}}
\title{Gauge field theory without groups}
\date{{\small November 12, 2020} }
\author{{D.\ Canarutto} \\[6pt]
{\small\it Dipartimento di Matematica e Informatica ``U.~Dini'', }\\
{\small\it Via S. Marta 3, 50139 Firenze, Italia}\\
{\small email:~{\tt canarutto@unifi.it}}\\
{\small {\tt http://www.dma.unifi.it/\char126canarutto}}}
\begin{document}
\maketitle \thispagestyle{empty}
\begin{abstract}\noindent
Non-standard topics underlying a partly original approach
to gauge field theory are concisely introduced,
expressing ideas that were broached in several papers and, eventually,
exposed in an organized form in a recently published book~\cite{Canarutto2020}.
By proposing a change of perspective
about the roles and relative importance of several notions,
this approach seeks to obtain an overall clarification of foundational matters.
In particular, by consistently relying on natural differential geometry,
the role of groups is shown to be downgradable to secondary.
\end{abstract}

\bigbreak\noindent
2010 MSC:
53B05, 
58A32, 
70Sxx, 
81R25, 
81R40. 
81V10, 
81V15, 
81T13, 
81T20, 
83C05. 

\smallbreak\noindent
Keywords: gauge field theory, natural geometric language, 2-spinors,
tetrad-affine gravity, Fr\"olicher smoothness, quantum bundles, quantum fields, quantum particles.

\tableofcontents
\newpage

\subsection{Introduction}

It is a fairly common opinion among mathematicians,
also shared by some theoretical physicists,
that the mathematical foundations of particle physics and quantum field theory
are somewhat fuzzy and unclear.\footnote{ 
`All too often in physics familiarity is a substitute for understanding',
Y.\ Choquet-Bruhat and C.\ DeWitt-Morette~\cite{ChoquetBruhatDeWittMorette}.} 
Moreover---partly as a consequence of the present scarcity
of sound experimental evidence---one has to deal with a plethora of theories
and approaches\footnote{
See e.g.\ R.\,Penrose~\cite{Penrose2016}.} 
that is certainly contrary to Occam's `razor principle'.

In recent years,
I attempted to clear up the field by revisiting several basic notions~[4--24].
Eventually, that work generated a fairly consistent view,
which was exposed in a new book~\cite{Canarutto2020}.
In this paper I sketch a few essential ideas,
sticking---as in the book---to concepts that are unquestionably relevant
to the foundations of present-day quantum physics.
I am not really interested in topics that, up to now,
seem to be only relevant to pure speculation.
Actually, I am convinced that choosing fewer notions and focusing on them,
rather than ever adding to them,
is the sole effective route to clarity,
and may even allow us to move beyond the present impasse in fundamental physics.

The main non-standard ideas considered, and briefly discussed in this presentation,
are all related, in one way or another,
to a preference for using a natural geometric language as far as possible.
Two-spinor algebra and two-spinor bundle geometry provide an integrated,
`minimal geometric data' approach to Einstein-Cartan-Maxwell-Dirac fields and,
more generally, to arbitrary-spin matter fields
coupled with gauge fields and tetrad-affine gravity.
Electroweak geometry and the ensuing field theory can be described
in a completely group-free approach, too.

The `configuration bundle' of gauge field theory,
whose sections are both matter and gauge fields,
can be introduced by intrinsic geometric constructions
without any reference to structure groups and matrix formalisms.
In this context, a `covariant differential' formulation of gauge field theory
conveniently replaces the usual jet bundle formulation,
and naturally extends to tetrad-affine gravity;
a first-order Lagrangian theory of fields with arbitrary spin is also exhibited.
Lie derivative of all fields, including the tetrad and the spinor connection,
are introduced, and lead to the notions of deformed field theory and energy tensors.

All the above concepts can be seamlessly extended to quantum bundles and quantum fields,
which are introduced by precise constructions based on Fr\"olicher's notion of smoothness.
Various aspects of QFT are exposed in a language that may appease a mathematically oriented reader
who is not satisfied with usual presentations of this matter
(in that case, I'd advise her to get the book).

Finally, I propose an approach to quantum particle physics
which is based on quantum geometry but, actually, bypasses the use of quantum fields.

One salient characteristic of the proposed approach is its consistent reliance on natural,
intrinsic geometric language,
with hardly any reference to structure groups.
Although these are still present in the background---as groups of automorphisms
of the considered geometric structures---I maintain that their role
can be quite conveniently regarded as auxiliary rather than primary.
The fact that this is actually possible without losing anything essential
will come as a surprise to many theoretical physicists,
and deserves a thorough discussion.


\subsection{Natural geometry and the role of groups}
\label{s:Natural geometry and the role of groups}

For historical reasons,
the standard approach to gauge field theory exploits the notion
of a fixed \emph{structure group}
via principal bundles and vector bundles associated with them.
Though the abstract notion of a vector space---possibly endowed
with some further algebraic structure---is not recent,
many still treat vectors as elements of $\RR^n$ (or $\CC^n$)
that `transform' according to a certain law.
Indeed, when a basis is chosen one gets such a representation,
and a vector's components in different bases are related by a transformation
that is an element of a group of matrices.
More precisely, the set $\mathrm{B}\V$ of all bases
of a vector space, $\V$\!, turns out to
have the structure of a \emph{group-affine} space,\footnote{
The straightforward formal definition is analogous to the usual
definition of affine space modelled on a vector space.} 
any two elements being related by an element of a group of matrices
which is isomorphic to the group \txtm{\Aut\!\V\subset\End\!\V}
of all automorphisms of $\V$ (constituted by all invertible endomorphisms).
However, there is no distinguished isomorphism between these groups;
rather, an arbitrary group isomorphism is singled out by the choice of a basis.

Usually one deals with vector spaces that are endowed with further
algebraic structure (for example, a scalar product).
This selects the subgroup
\txtm{\Gb\subset\Aut\!\V} of all automorphisms preserving that structure,
and the subset of $\mathrm{B}\V$ constituted by all `special' bases
(for example, orthonormal bases),
whence one may recover the usual matrix groups.
The actual use of matrices, however, is basically about calculations,
while natural geometric notions correspond to a higher level of abstraction
and are best suited to understanding the fundamental concepts.
Accordingly, I feel justified in seeing the role
of matrices and matrix groups as secondary.

Some further clarification will be opportune in the discussion of the meaning of `intrinsic'.
Generally speaking, a differential geometric notion is called \emph{intrinsic}
if it corresponds to a well-defined geometric object
that can be directly characterized in terms of natural operations
allowed by the underlying geometric structure.
Thus `intrinsic' may be seen as roughly equivalent
to the basic acceptation of `covariant' in a broad sense.
In the mathematics literature, several intrinsic objects and operations
are often denoted in coordinate-free form,
through symbols that give up indices and matrices.
This economy of language is not just convenient shorthand:
it corresponds to true mathematical abstraction.

Now we should distinguish between algebraic notions,
related to the fibre structure of the involved bundles,
and actual differential geometric notions arising from the basic concepts
of (fibred) manifold and jet space.
The former, including all notions derived from tensor products
and scalar products of any type,
can be viewed as truly intrinsic from the start:
indeed, they can be introduced, and their fundamental properties established,
while making no use of bases and matrices.
The concept of manifold, however, originates from that of cocycle
(a family of compatible charts covering an assigned topological space).
Though charts need not be valued into $\RR^n$
(an affine space is sufficient),
natural notions in this context are exactly defined by the requirement
that their chart expressions are `invariant'
with respect to transformations between compatible charts.
However, once the fundamental such notions have been introduced
and their main properties have been established by checking invariance,
such checking need not be performed at successive stages of development:
definitions and demonstrations only use the natural properties;
if these are correctly implemented,
then invariance is automatic and it is no longer necessary to prove it.

Thus in a general field theory the term `natural language'
actually refers to a superposition of two different concepts.
In the standard physics literature, group invariance 
is mainly tied to the algebraic structure of bundle fibres in gauge field theory,
while it is usually less relevant in expositions of General Relativity.
I maintain that, in both senses,
the economy of language and deeper understanding
provided by intrinsic language---by comparison with
a language based on matrices and transformation rules---is well worth the effort.
A theoretical physicist might regard the familiar group approach as more succinct,
since it enables introducing a theory by specifying a group.
Such evaluation, however, depends on the assumed background;
if the whole machinery of principal bundles is needed,
that constitutes a rather substantial assumption!
Furthermore, this fixed general recipe may eventually make `thinking differently' harder.

That said, it is worthwhile to point out that thoroughly giving up coordinate expressions,
although possible in principle,
would impose a notation too awkward to be useful in practice;\footnote{ 
A computer code effectively treating tensor algebra
in a fully intrinsic form is certainly feasible,
but its output would be hard for us to interpret.
This type of difficulty, however, also applies to
the internal form of even simple expressions
in a symbolic language like Wolfram's Mathematica.} 
in particular, one needs to deal with tensor products with several factors
and many possible different contractions, symmetrizations, and antisymmetrizations.
Now, the index formalism was introduced exactly for dealing with such issues;
it closely reflects the intricacies arising in tensor algebra and its extensions.
If a coordinate expression is intrinsically well constructed
as a representation of natural operations,
then checking its `covariance' with respect to the appropriate group of transformations
is an inessential exercise.
The best route is then a middle one,
in which coordinate-free expressions are reserved for the most important
objects and operations, while coordinates and indices
retain their role as efficient means of performing
computations and demonstrations.\footnote{ 
In Ch.\,2 (Vol.\,I) of their well-known monograph,
\emph{Spinors and space-time}~\cite{PenroseRindler88},
Penrose and Rindler discuss an `Abstract index notation'
aimed at resolving this duplicity of the index formalism;
subsequently that notation is not extensively used, however.
In the appendix of the same book they also discuss a diagrammatic approach
to tensor algebra, noting that, in practice, its actual use is limited.} 
Another way to see this is the following:
tensor products represent multilinear maps; each factor can be seen as a `slot' that functions either as input or output, and for efficient calculations the slots must be labelled.
The index formalism provides a convenient labelling system.
Thus an indexed expression need not be regarded
as representing a matrix of components.

Also note that, in most cases,
one need not be actually involved with the global topological aspects of the
bundles under consideration;
just assuming the topology allows all the needed geometric constructions is enough.
Namely, one deals with \emph{local}, \emph{intrinsic} differential geometry;
the two italicized adjectives are fully compatible,
although `local' is often somewhat misleadingly
used in the sense of `coordinate-dependent'.

\subsection{Configuration bundle of classical gauge field theory}
\label{s:Configuration bundle of classical gauge field theory}

In order to describe gauge fields in intrinsic terms, let us start
from the observation that the vector space \txtm{\End\!\V\cong\V{\otimes}\V^*}
of all endomorphisms of a vector space, $\V$\!, has a natural structure of Lie algebra
determined by the ordinary commutator,
and any specialized algebraic structure yields the Lie subalgebra
\txtm{\Lie\subset\End\!\V} of its `infinitesimal symmetries'.
If the algebraic structure is a real or Hermitian scalar product,
then $\Lie$ is constituted, respectively, by all antisymmetric endomorphisms
and all anti-Hermitian endomorphisms
(with respect to the scalar product itself).
As for spinors, and everything that is associated with them,
the only fundamental algebraic structure that is actually needed
is that of a two-dimensional complex vector space
(\emph{two-spinor space}, see~\Sec\ref{s:Two-spinor geometry}).
In all cases, any references to groups can be thoroughly dropped.

In field theories the above notions can be exploited
in terms of vector bundles smoothly endowed with some fibre structure.
Locally, one may recover the traditional principal bundle approach
in terms of the bundle of frames that are special with respect to the
considered fibre structure (a group-affine bundle).
Lie-algebra bundles, on the other hand,
have a definitely relevant role in gauge field theory---though
the matricial formalism can be still regarded as marginal.

In a typical gauge field theory a \emph{matter field}
can be described as a section, \txtm{\phi:\M\to\E}, of a vector bundle
(usually, the base manifold $\M$ is the spacetime manifold),
while a \emph{gauge field} is a special connection.
Thus a gauge field cannot be described as a section of some vector bundle.
Indeed, an arbitrary connection is a section
\txtm{\E\to\JO\E} of the first-jet prolongation bundle,
so it cannot even be described as a section
of any finite-dimensional bundle over $\M$. 
However, we may choose to select those connections which preserve
the algebraic structure of a vector bundle,
called \emph{linear connections}.\footnote{
The `affine' label often attached to connections is related to the fact that,
even when \txtm{\E\onto\M} is a generic fibred manifold
with no algebraic fibre structure, the bundle
\txtm{\JO\E\onto\E} turns out to be naturally affine.
This fact should not generate confusion with labels used
to identify connections selected by specific algebraic structures.} 
These can indeed be regarded as sections of a finite-dimensional bundle,
namely the affine sub-bundle\footnote{
If \txtm{\E\onto\M} is a vector bundle,
then \txtm{\JO\E\onto\M} is a vector bundle, too.} 
\txtm{\LO\!\CO\E\subset\JE\ten{\M}\E^*} over $\M$
which projects onto the identity section \txtm{\Id\E:\M\to\E\ten{\M}\E^*};
its associated vector bundle is
$$\TS\M\ten{\M}\End\!\E\equiv\TS\M\ten{\M}\E\ten{\M}\E^*\onto\M~.$$

Any further algebraic fibre structure of \txtm{\E\onto\M} beyond linearity
selects a Lie-algebra sub-bundle, \txtm{\Lie\subset\End\!\E}\,;
connections that make that algebraic structure \emph{covariantly constant}
can be characterized as sections of an
affine sub-bundle, \txtm{\K\subset\LO\!\CO\E},
whose associated vector bundle is \txtm{\TS\M\ten{\M}\!\Lie\onto\M};
gauge fields are exactly sections \txtm{\kg:\M\to\K}. Groups have no role in these definitions.

Differential operations related to gauge fields can be expressed in terms of the
Fr\"olicher--Nijenhuis bracket of vector-valued forms.
If \txtm{\zeta:\M\to\weu{r}\ten{\M}\TM} and \txtm{\xi:\M\to\weu{s}\ten{\M}\TM}
are tangent-valued forms on a generic manifold, $\M$,
then their Fr\"olicher--Nijenhuis bracket is a tangent-valued form\footnote{
Characterized, via linearity, by the rule
\smallbreak\noindent
\txtm{\fnb{\la\tn u,\mu\tn v}=\la\we\mu\tn[u,v]+\la\we(\LO[u]\mu)\tn v-(\LO[v]\la)\we\mu\tn u
+(-1)^r(v|\la)\we\dO\mu\tn u +(-1)^r\dO\la\we(u|\mu)\tn v}\,,
\smallbreak\noindent
where \txtm{\la:\M\to\weu{r}\TS\M}\,, \txtm{\mu:\M\to\weu{s}\TS\M}\,, \txtm{u,v:\M\to\TM}\,,
and $[u,v]$ is the Lie bracket of $u$ and $v$\,.} 
\txtm{\fnb{\zeta,\xi}:\M\to\weu{r+s}\ten{\M}\TM}\,.
After replacing the generic manifold $\M$ with the total manifold of a fibred manifold, $\E$\,,
one considers special cases such as vertical-valued forms \txtm{\E\to\weu{r}\TS\E\ten{\E}\VE}
and horizontal forms \txtm{\E\to\weu{r}\TS\M\ten{\E}\TE}\,.
Because of the natural inclusion \txtm{\JE\subset\TS\M\ten{\E}\TE}\,,
a connection can be included in the latter.
Moreover, if \txtm{\E\onto\M} is a vector bundle, then we have the natural isomorphism
\txtm{\VE\cong\E\cart{\M}\E}\,,
so that a matter field can be viewed as a vertical valued zero-form.
Hence we get the \emph{covariant differentials}
\[ \dO[\kg]\phi\equiv\fnb{\kg,\phi}:\M\to\TS\M\ten{\M}\E~,\qquad
\dO[\kg]\kg\equiv\fnb{\kg,\kg}:\E\to\weu2\TS\M\ten{\M}\E~,\]
which can be regarded, via obvious identifications,
as the covariant derivative \txtm{\nabla[\kg]\phi} of the matter field
and minus the curvature tensor of the gauge field.
Furthermore, the latter is also linear
(because of the linearity of the connection $\kg$),
so that it can be regarded as a section
\[ \uprho\equiv-\dO[\kg]\kg:\M\to\weu2\TS\M\ten{\M}\Lie\subset
\weu2\TS\M\ten{\M}\End\E~.\]

Therefore, the `configuration bundle' of a generic gauge field theory is \txtm{\E\cart\M\K\onto\M}\,,
a field is a couple \txtm{(\phi,\kg):\M\to\E\cart\M\K}\,,
and its \emph{covariant prolongation} is
\begin{align*}
\dO(\phi,\kg)&\equiv(\phi,\kg,\dO[\kg]\phi,\dO[\kg]\kg):\M\to\dO\bigl(\E\cart{\M}\K\bigr)
\equiv
\\[6pt]
&\hspace{140pt}
\equiv\E\cart{\M}\K\cart{\M}
\bigl(\TS\M\ten{\M}\E\bigr)\cart{\M}\bigl(\weu2\TS\M\ten{\M}\Lie\bigr)
\end{align*}
Of course, the above basic setting will have to be expanded and adapted to various special cases.
The principal bundle formalism, though familiar to many, is comparatively fairly intricate.
One may argue about the convenience of abandoning it.
The short answer to this is that the effort of taking a step further in abstraction
does pay off, by providing an optimized language
that helps to prevent confusion or diversion by spurious ideas.
On a related note, let me remark that in the literature one finds
considerable confusion between the notion of observer
and the notions of frames and coordinates,
with the adjective `covariant' often being used in the sense of
`independent of the observer'.

The notion of `gauge fixing' has various aspects, which I am not going to discuss here.
Just to touch one important point: the quantum version of a classical field
requires that the field be a section of a vector bundle,
while a gauge field is a section of an affine bundle.
A way out of this difficulty consists of \emph{fixing a gauge},
namely a local `background' curvature-free connection, $\kg_\0$\,,
to use as a \emph{reference}:
now any gauge field $\kg$ can be represented
as the difference \txtm{\kg\,{-}\,\kg_\0}\,, a tensor field.
Usually $\kg_\0$ is seen as associated with the choice of a local frame,
though there is no obligation with regard to this.


\subsection{Covariant differential and Lagrangian field theory}
\label{s:Covariant differential and Lagrangian field theory}

The appropriate geometric language for a general Lagrangian field theory of arbitrary order
is that of jet bundles.
An actual gauge field theory related to particle physics, however, is more specific:
its Lagrangian is of the first order, and depends on the fields' derivatives
only through their covariant differentials (\emph{Utiyama principle}).
Furthermore, the Lagrangian depends on the fields and their differentials \emph{polynomially}
(no arbitrary functional dependence).
Hence, the derivatives of the Lagrangian with respect to the fields and their differentials
are actually simple \emph{algebraic} operations;
this fact is essential for a seamless extension to the quantized theory
(\Sec\ref{s:Quantum bundles and quantum fields}).

Thus the Lagrangian density can be expressed as a morphism
\[ \Lagr:\dO\bigl(\E\cart{\M}\K\bigr)\to\weu{m}\TS\M~,\qquad m\equiv\dim\M~,\]
so that
\[ \Lagr[\phi,\kg]\equiv\Lagr\comp\dO(\phi,\kg):\M\to\weu{m}\TS\M \]
is an ordinary density on $\M$.
By analogy with the momentum form associated with the Lagrangian density in the usual formulation,
one introduces the sections 
\begin{align*}
\Pi^{\zero}&\equiv\frac{\de\Lagr[\phi,\kg]}{\de\phi}:\M\to\weu{m}\TS\M\tn\E^*~,
\\[6pt]
\Pi^{\one}&\equiv\frac{\de\Lagr[\phi,\kg]}{\de(\dO[\kg]\phi)}:\M\to\weu{m-1}\TS\M\tn\E^*~,
\\[6pt]
\Pi^{\two}&\equiv\frac{\de\Lagr[\phi,\kg]}{\de(\dO[\kg]\kg)}:\M\to\weu{m-2}\TS\M\tn\Lie^*~,
\end{align*}
where the derivatives are, in practice, algebraic operations, as observed above.

The reader will note how $\Pi^{\zero}$, $\Pi^{\one}$, and $\Pi^{\two}$
can be regarded as vector-valued forms.
Accordingly, it can be proved that the Euler--Lagrange equations
for the couple $(\phi,\kg)$ can be cast in the form
\[\begin{cases}
\Pi^{\zero}-\dO[\kg]\Pi^{\one}=0~,
\\[8pt]
\lfr^*(\Pi^{\one}\tn\phi)-\dO[\kg]\Pi^{\two}=0~,
\end{cases}\]
where \txtm{\,\lfr:\Lie\into\End\!\E\cong\E{\otimes}\E^*}
denotes the natural inclusion, and
\[\lfr^*:\End\!\E^*\cong\E^*{\otimes}\E\to\Lie^* \]
is its transpose morphism.

The above scheme turns out to be naturally extendable to more intricate situations
in which one has several matter and gauge sector, variously interacting.
Furthermore, tetrad-affine gravity can also be treated in this way
(\Sec\ref{s:Two-spinor geometry}).

A further remark: in many cases, a matter field is actually a couple of fields
\[ (\phi,\phi^*):\M\to\E\cart\M\E^*~.\]
When a Hermitian fibre metric is assigned,
the field equation often admit solutions in which $\phi$ and $\phi^*$
are mutually conjugate transpose fields.
Nevertheless, they can (and, for clarity, should) always be viewed as mutually independent.


\subsection{Two-spinor geometry}
\label{s:Two-spinor geometry}
The `minimal geometric data' approach broached in this section
essentially consists of a complex bundle, \txtm{\S\onto\M}\,, with two-dimensional fibres,
without any further assumptions.
I will first sketch the fibre-algebraic aspect of this approach,
in terms of a two-dimensional complex vector space, $\S$\,;
the considered construction are closely related
to the Penrose-Rindler formalism~\cite{PenroseRindler84,PenroseRindler88},
though there are differences that I will not examine in detail here.
Note that, although the two-spinor index formalism is useful in many computations,
everything can be expressed intrinsically.

Let us start from the observation that any finite-dimensional complex vector space, $\V$,
has a dual space, $\Vl$, and \emph{anti-dual} space, $\Val$
(consisting of all anti-linear functions \txtm{\V\to\CC}),
and a \emph{conjugate space}, \txtm{\Vc\equiv\V^{\lin\alin}};
then one also gets the natural identifications \txtm{\Val\cong\Va},
\txtm{\Vc\cong\V^{\alin\lin}}.
If \txtm{\la\in\Vl}, then complex conjugation yields \txtm{\bar\la\in\Va}
by the rule \txtm{\bar\la(v)\equiv\overline{\la(v)}},
namely anti-isomorphisms \txtm{\Vl\leftrightarrow\Va} and, similarly, \txtm{\V\leftrightarrow\Vc}.
Thus, one gets the real-linear involution (\emph{Hermitian transposition})
\[ \dagger:\V\tn\Vc\to\V\tn\Vc:u\tn\bar v\mapsto(u\tn\bar v)^\dagger\equiv v\tn\bar u~, \]
extended by linearity;
this, in turn, determines the splitting
\[ \V\tn\Vc=\HO(\V\tn\Vc)\oplus\iO\,\HO(\V\tn\Vc) \]
into the eigenspaces corresponding to eigenvalues $\pm1$ of $\dagger$,
called the \emph{Hermitian} and \emph{anti-Hermitian} subspaces.
When applied to the case of a two-dimensional complex vector space, $\S$\,,
these constructions generate a rich algebraic structure.

\subsubsection*{\emph{Complex symplectic structure and the space of length units}}

The antisymmetric vector subspace \txtm{\weu2\S\subset\S\tn\S} is one-dimensional.
Thus, the Hermitian subspace of \txtm{\weu2\S\tn\weu2\Sc} is a one-dimensional
real vector space, which also turns out to have a natural orientation:
the positive subspace of all elements of the type $w\tn\bar w$\,, with \txtm{w\in\weu2\S}\,,
is denoted by \txtm{\LL^2\equiv\LL\tn\LL}\,;
the positive space $\LL$ will be identified
with the semi-vector space of length units.\footnote{
For a comprehensive account of the geometry of semi-vector spaces,
and its application to a rigorous mathematical treatment of physical scales,
see~\cite{JanyskaModugnoVitolo10}.
A not-too-short introduction to this topic can also be found
in the aforementioned book~\cite{Canarutto2020}.} 

The rational power of a unit space is a well-defined unit space.\footnote{
If $\UU$ is a unit space, then we set \txtm{\UU^p\equiv\UU\otimes\dots\otimes\UU}
(\txtm{p\in\NN} factors)
The $r$-root of a unit space $\UU$ (\txtm{r\in\NN}) is a unit space, $\UU^{1/r}$,
which is charactyerized, up to isomorphism, by \txtm{(\UU^{1/r})^r=\UU}\,.
Moreover \txtm{\UU^{-1}\equiv\UU^*} (dual space),
so that \txtm{\UU^{\pm p/r}} is a well-defined unit space for \txtm{p\neq0}\,.} 
The two-dimensional complex vector space \txtm{\U\equiv\LL^{1/2}\tn\S} 
of `conformally invariant' two-spinors has an important role.
In particular, there is a distinguished Hermtian metric on $\weu2\U$\,,
yielding, \emph{up to a phase factor},
a unque `normalized complex symplectic form' \txtm{\upepsilon\in\weu2\Ul}.
This yields the isomorphisms \txtm{\upepsilon^\flat:\U\to\Ul},
defined by \txtm{\bang{\upepsilon^\flat(u),v}\equiv\upepsilon(u,v)}\,,
and \txtm{\upepsilon^\#\equiv-(\upepsilon^\flat)^{-1}:\Ul\to\U}.

\subsubsection*{\emph{Two-spinor generated Minkowski space}}

Since \txtm{\upepsilon\in\weu2\Ul} is unique up to a phase factor,
\txtm{\geu\equiv\upepsilon\tn\bar\upepsilon\in\weu2\U\tn\weu2\Uc} is a natural object,
which can be regarded as the bilinear form on $\U\tn\Uc$ characterized by the rule
\[ \geu(u\tn\bar v,u'\tn\bar v')\equiv\upepsilon(u,u')\,\bar\upepsilon(\bar v,\bar v')~.\]
Moreover, the Hermitian subspace \txtm{H\subset\U\tn\Uc}
is a 4-dimensional real vector space, and the restriction of $\geu$ to $\H$
turns out to be a Lorentz metric.\footnote{
In particular, any basis of $\U$ yields a $\geu$-orthonormal basis of $\H$,
expressed in terms of the former via Pauli matrices.} 

Isotropic vectors in $\H$ are of the form $\pm u\tn\bar u$\,, with \txtm{u\in\U}.
Thus one gets a natural time orientation of $\H$.


%
\subsubsection*{\emph{Dirac spinors}}
Let \txtm{\W\equiv\U\oplus\Uc}.
Consider the linear map \txtm{\dga:\U\tn\Uc\to\End\W}, characterized by
\[ \dga[p\tn\bar q](u,\bar\lambda)\equiv
\sqrt2\,\bigl(\bang{\bar\lambda,\bar q}\,p\,,\,\upepsilon(p,u)\,\bar\upepsilon^\flat(\bar q)\bigr)~.\]
Then, the restriction of $\dga$ to the Minkowski space \txtm{\H\subset\U\tn\Uc}
turns out to be a Clifford map (the \emph{Dirac map}).
The fact that $\upepsilon$ is unique up to a phase factor makes $\dga$ natural.

Thus, we are led to view the 4-dimensional complex space $\W$ as the space of Dirac spinors.
It is naturally endowed with a further structure, namely the isomorphism
\[ \Wc\to\Wl:(\bar u,\lambda)\mapsto(\lambda,\bar u)~;\]
this is associated with a Hermitian metric, $\kD$\,,
which has the signature \txtm{({+}{+}{-}{-})}\,.
If \txtm{\psi\in\W}, then \txtm{\bar\psi\in\Wc} can be regarded as an element in $\Wl$,
the \emph{Dirac adjont} of $\psi$\,.
It is not difficult to recover all the associated notions and identities
found in standard expositions of Dirac spinors;
in particular, \txtm{\dga[y]\in\End\W} is a $\kD$-Hermitian endomorphism
for any \txtm{y\in\H}.

On the other hand, the \emph{charge-conjugation} anti-isomorphism
\[ \mathcal{C}_\upepsilon:\W\to\W:(u,\bar\lambda)\mapsto
\bigl(\upepsilon^\#(\lambda),\bar\upepsilon^\flat(\bar u)\bigr) \]
does depend on an overall phase factor.


%
\subsubsection*{\emph{Observer-related structures}}

The above sketched spinor algebra setting does not assume distinguished positive Hermitian metrics
on the two-spinor space $\U$ or the 4-spinor space $\W$.
Indeed, a positive Hermitian tensor \txtm{\heu\in\Ua\tn\Ul} can be naturally identified
with a future-oriented, timelike covector in \txtm{\H^*}.
The assignment of such an object determines the anti-isomorphism
\txtm{\heu^\flat:\U\to\Ul:u\mapsto\heu(\bar u,\_)}\,;
in turn, this yields the anti-isomorphism \txtm{\W\to\Wl}
which is usually denoted\footnote{
Note that the Dirac adjunction \txtm{\psi\mapsto\bar\psi},
which is observer-independent, is often introduced as the combination
\txtm{\bar\psi=\psi^\dagger\dga_0}\,,
that is, in terms of two observer dependent operations.} 
by \txtm{\psi\mapsto\psi^\dagger}.

A $\geu$-normalized, future-oriented, timelike vector \txtm{\tau\in\H}
is called an \emph{observer}.
From the Dirac algebra identity \txtm{\dga[\tau]\comp\dga[\tau]=\id} one sees that
an observer determines the splitting \txtm{\W=\W^+_{\!\tau}\oplus\W^-_{\!\tau}}
into eigenspaces of $\dga[\tau]$ with eigenvalues $\pm1$
(which is related to the distinction between electrons and positrons).
Furthermore, an observer yields the \emph{parity} and \emph{time-reversal} operators,
and the \emph{spin operators}.
The whole standard spinorial machinery can be then recovered in a very direct
and natural way.


%
\subsubsection*{\emph{Two-spinor bundle and spacetime geometry}}

A two-spinor bundle \txtm{\S\onto\M} determines vector bundles
\txtm{\U\onto\M}, \txtm{\H\onto\M}, \txtm{\W\onto\M}, and so on;
their fibers are smoothly endowed with the previously sketched algebraic structures.

A linear connection \txtm{\cs} of \txtm{\S\onto\M} yields
linear connections of the induced bundles;
in particular, the induced connection $\Gamma$ of \txtm{\H\onto\M}
turns out to be metric, namely one has \txtm{\nabla[\Gamma]\geu=0}\,.

The above considerations hold independently of the base manifold $\M$\,;
now, assuming that $\M$ is 4-dimensional,
one defines a \emph{soldering form} (or \emph{tetrad})\footnote{
The term `tetrad' is a convenient shorthand,
but is somewhat misleading as it was introduced to indicate
an orthonormal spacetime frame---and it is usually still intended in that way.
A soldering form, on the other hand, is a fully intrinsic notion.
If one chooses an orthonormal frame of \txtm{\H\onto\M},
then indeed recovers a `tetrad formalism' which is similar to the usual one.} 
as a fibred isomorphism
\[ \uptheta:\TM\to\LL\tn\H\subset\S\tn\Sc~.\]
A tetrad can also be regarded as a section\footnote{
If no confusion arises, fibred tensor products
will be denoted as plain tensor products.} 
\[ \uptheta:\M\to\LL\tn\H\tn\TS\M~.\]

The requirement that $\uptheta$ be non-degenerate, namely an isomorphism,
is needed to recover its standard physical interpretation;
but it is not actually needed to use it as a `sector' in a
fledged field theory (\Sec\ref{s:Gauge field theories and tetrad-affine gravity}).
A non-degenerate soldering form determines a Lorentz metric on $\M$
by `translating' the Lorentz metric $\geu$ of $\H$.
Moreover, the soldering form together with a spinor connection
yield a linear connection, $\Gamma$, of the tangent bundle \txtm{\TM\onto\M},
which turns out to be metric but, in general, has non-vanishing torsion.
Although $\geu$ also yields the Levi-Civita connection,
coupling with spinor fields is a source for torsion.
Actually, the induced spacetime connection is just a byproduct,
not a fundamental field;
instead, the gravitational field the \emph{tetrad-affine} setting
is represented by the couple $(\uptheta,\cs)$.
One gets the identities
\[ \uptheta\pint\Teu=\fnb{\Gamma,\uptheta}~,\quad
0=\fnb{\uptheta,\fnb{\Gamma,\Gamma}}+\fnb{\Gamma,\fnb{\Gamma,\uptheta}}
+\fnb{\Gamma,\fnb{\uptheta,\Gamma}}~,\quad
0=\fnb{\Gamma,\fnb{\Gamma,\Gamma}}\equiv-\fnb{\Gamma,\Reu}~,\]
where $\Teu$ and $\Reu$ are the torsion and the Riemann tensor of the induced connection $\Gamma$.
The last two identities are the first and second Bianchi identities, respectively.

If \txtm{\psi:\M\to\W}, then \txtm{\nabla\psi\equiv\nabla[\cs]\psi:\M\to\TS\M\tn\W};
by performing a few natural contractions in \txtm{\cev\uptheta\tn\dga\tn\nabla\psi}\,,
where $\cev\uptheta$ is the inverse isomorphism of $\uptheta$\,,
one gets a section \[\nasl\psi:\M\to\LL\!^{-1}\tn\W~.\]
This construction defines the \emph{Dirac operator} $\nasl$.
If $\uptheta$ is degenerate, then a similar construction can be performed
by replacing $\cev\uptheta$ with \txtm{\weu3\uptheta:\M\to\LL\!^3\tn\weu3\TS\M\tn\weu3\H},
thus obtaining a section
\[ \breve\nasl\psi:\M\to\LL\!^3\tn\W\tn\weu4\TS\M \]
which fulfils \txtm{\breve\nasl\psi=\nasl\psi\tn\upeta}
($\upeta$ is the metric volume form).

\comment{
Two further constructions play a relevant role.
The third exterior power of $\uptheta$ is the section
\txtm{\weu3\uptheta:\M\to\LL\!^3\tn\weu3\TS\M\tn\weu3\H},
which by contraction with the natural fibre volume form $\upeta$ of $\H$ yields the
\emph{complementary soldering form} (or \emph{co-tetrad})
\[ \breve\uptheta:\M\to\LL\!^3\tn\weu3\TS\M\tn\H^*~.\]
If $\uptheta$ is non-degenerate (hence it has an inverse, $\cev\uptheta$)
then \txtm{\upeta\equiv\uptheta^*\upeta}
is the induced volume form on $\M$,
and the identity \txtm{\breve\uptheta=\cev\uptheta\tn\upeta} holds.

Finally, if \txtm{\psi:\M\to\W}, then the contraction of
\txtm{\nabla[\cs]\psi:\M\to\TS\M\tn\W} with $\breve\uptheta$ yields a section
\[ \breve\nasl\psi:\M\to\LL\!^3\tn\W\tn\weu4\TS\M~,\]
}

\subsection{Gauge field theories and tetrad-affine gravity}
\label{s:Gauge field theories and tetrad-affine gravity}

In the context sketched in~\Sec\ref{s:Two-spinor geometry},
one gets a  `minimal geometric data' approach
to Einstein--Cartan--Maxwell--Dirac fields:
all the required underlying structures are derived by natural geometric constructions
from a complex bundle \txtm{\S\onto\M} with two-dimensional fibres.
One has the induced bundles
\txtm{\LL\onto\M}, \txtm{\U\onto\M}, \txtm{\H\onto\M}, \txtm{\W\onto\M},
with their natural fibre structures,
and considers the following fields:
\begin{mybulist}
\item
the soldering form \txtm{\uptheta:\M\to\LL\tn\TS\M\tn\H}
\item
the two-spinor connection
\txtm{\cs:\M\to\LO\!\CO\U} 
\item
the Dirac field \txtm{\psi\equiv(u,\bar\lambda):\M\to\LL\!^{-3/2}\tn\W}
\item
the dual Dirac field
\txtm{\bar\psi\equiv(\bar u,\lambda):\M\to\LL\!^{-3/2}\tn\Wc}
\item
the Maxwell field \txtm{{\st F}:\M\to\LL\!^{-2}\tn\weu2\H^*}\,.
\end{mybulist}

A further field should actually be considered, namely a \emph{dilaton} field \txtm{\lu:\M\to\LL}\,.
This is eliminated if we make just \emph{one} a-priori hypothesis
about the theory: the connection of \txtm{\LL\onto\M}
determined by the spinor connection is flat.
Then \txtm{\LL\onto\M} is trivial and, in practice,
one deals with a fixed space $\LL$ of unit lengths
(the notion of a coupling \emph{constant} now makes sense).
Note how the fields are \emph{scaled}, namely, tensorialized by powers of $\LL$
(this is true if one uses the \emph{natural} unit settings,
in which \txtm{\hbar=c=1}, otherwise one deals with further unit spaces).
In coordinate expressions, the scaling can be conveniently attributed to the field components,
and the tensor product by a scaling factor is indicated by simple juxtaposition;
in this way one gets notations close to the usual one.

It should be noticed that the spinor connection
yields a Hermitian connection ${\st Y}$ on \txtm{\weu2\U\onto\M},
which can be regarded as the electromagnetic potential:
its relation to the Maxwell field will be a consequence of the field equations.\footnote{
Thus, ${\st F}$ and ${\st Y}$ are independent fields.
} 

The Lagrangian density can be now expressed as
\txtm{\Lambda=\Lambda\Dir+\Lambda\emag+\Lambda\grav}\,,
where
\begin{align*}
\Lambda\Dir&=\bigl(\tfrac\iO2\,(\bang{\bar\psi,\nasl\psi}-\bang{\nasl\bar\psi,\psi})
-m\,\bang{\bar\psi,\nasl\psi}\bigr)\,\upeta~,\qquad m\in\LL\!^{-1}~,
\\[6pt]
\Lambda\emag&=\tfrac14\,\Bigl\langle\geu^\#\!({\st F})\,,\,
\bang{\weu2\uptheta,\dO{\st Y}}+{\st F}\,\upeta\Bigr\rangle~,
\\[6pt]
\Lambda\grav&=\tfrac1{4\st\mathbb{G}}\,\bang{\weu2\uptheta,\Reu}~,
\end{align*}
where $\st\mathbb{G}$ is Newton's gravitational constant.
The field equation can now be straightforwardly derived
with the procedure outlined in~\Sec\ref{s:Covariant differential and Lagrangian field theory}.
Note that, in the `gravitational sector',
$\uptheta$ formally plays the role of the matter field and $\cs$ plays the role of the gauge field.
One gets: the Dirac equation with torsion;
the relation \txtm{{\st F}=2\,\dO{\st Y}}, and the second Maxwell equation with the Dirac current;
the Einstein gravitational equation, and a further `torsion equation'
involving the torsion and the Dirac field.

The above scheme can be seamlessly extended to include matter fields of arbitrary spin,
and gauge fields valued into non-trivial Lie algebras.
The details of such extensions lie outside the scope of this summary,
but let us briefly look at a few points.
\begin{mybulist}
\item
In general, the gauge Lagrangian can be expressed similarly to the e.m.\ case.
Contraction in the fibres of the Lie algebra bundle \txtm{\Lie\onto\M}
are performed via the natural scalar product
\txtm{(\xi,\zeta)\mapsto\mathrm{Tr}(\xi^\dagger\comp\zeta)},
where \txtm{\xi\mapsto\xi^\dagger} denotes adjunction with respect to the
scalar product which determines \txtm{\Lie\subset\End\E}.
\item
The matter field can be a section of a bundle of the type \txtm{\E\ten\M\Z\onto\M},
where \txtm{\Z\onto\M} is generic `internal spin' bundle
constructed  from any fibred tensor products
and direct products (over $\M$) of $\U$, $\Ul$, $\Uc$, $\Ua$.
The vector bundle \txtm{\E\onto\M}, instead,
is \emph{unsoldered} from spacetime:
its sections are \emph{spin-zero} fields possessing further `internal degrees of freedom'.
\item
For all types of matter fields one has the \emph{Klein--Gordon} Lagrangian
\[ \Lambda\spec{KG}\equiv
\tfrac12(\bang{\geu^\#\nabla\!\phi^*\tn\nabla\!\phi}-m^2\,\bang{\phi^*\tn\phi})\,\upeta~,\]
where $\geu^\#$ is the inverse (`contravariant') spacetime metric
and angle brackets indicate that all possible contractions are taken.
\item
For fields of arbitrary spin, a first-order generalization of the Dirac equation
can be introduced, though the procedure is somewhat more intricate.
The action of the Dirac map can be extended to a bundle $\Z$ of the above said type,
but in general $Z$ is not closed with respect to such action,
which generates \emph{ghost sectors} \txtm{\Z',\Z'',\dots}
forming a closed sequence \txtm{\Z\to\Z'\to\Z''\to\dots\to\Z}.
A natural Lagrangian yielding a first order generalized Dirac equation can be exhibited;
in flat spacetime, a plain wave solution in the main sector $\Z$ determines
the solution in the ghost sectors.\footnote{
Besides the aforementioned book, details about this subject can be found
in a dedicated paper~\cite{Canarutto16f}.} 
\item
The electroweak theory can be formulated within two-spinor geometry,
without dealing with structure groups,
by adding one main ingredient:
the \emph{isospin bundle} \txtm{\I\onto\M},
a Hermitian bundle with two-dimensional complex fibres.
The fermion bundle is then assumed to be the fibred direct product
of a \emph{right-handed} and \emph{left-handed} sectors,\footnote{
In an even more extended theory, the fermion bundle can be written as
\txtm{(\E\spec{R}\tn\U)\oplus(\E\spec{L}\tn\Ua)}\,.} 
that is
\[ \Y\equiv\YR\dir\M\YL\equiv\bigl(\weu2\I\ten{\M}\U\bigr)\dir\M\bigl(\I\ten\M\Ua\bigr)~.\]
The electroweak fields are then
\begin{mybulist}
\item
the fermion field
\txtm{\Psi\equiv\Psi\!\spec{R}\,{+}\,\Psi\!\spec{L}:
\M\to\LL\!^{-3/2}\tn(\YR\,{\oplus}\,\YL)} \smallbreak
\item
the anti-fermion field
\txtm{\bar\Psi:\M\to\LL\!^{-3/2}\tn(\YR\,{\oplus}\,\YL)^\lin} \smallbreak
\item
the gauge field
\txtm{{\st W}:\M\to\LL\!^{-1}\tn\H^*\tn\Lie\spec{L}} \smallbreak
\item
the Higgs field
\txtm{\phi:\M\to\LL\!^{-1}\tn\weu2\Ic\tn\I\cong\LL\!^{-1}\tn\I\tn\weu2\Il} \smallbreak
\item
the anti-Higgs field
\txtm{\bar\phi:\M\to\LL\!^{-1}\tn\weu2\I\tn\Ic\cong\LL\!^{-1}\tn\weu2\I\tn\Il}\,,
\end{mybulist}
where \txtm{\Lie\spec{L}\subset \I\tn\Il}
is the anti-Hermitian Lie-subalgebra bundle.
For any assigned gauge, ${\st W}$ yields a connection\footnote{
Here I shift from the view that a gauge field is a connection to the view that
a connection and a gauge field are actually strictly related but distinct notions,
with different roles.
This approach is thoroughly discussed in the book~\cite{Canarutto2020},
but such discussion lies outside the scope of this summary.} 
of \txtm{\YL\onto\M}.
The gauge field $\hat{\st W}$ in the right-handed sector
is naturally determined by ${\st W}$.

The full-fledged electroweak theory requires two further ingredients:
a fixed, background Higgs field $\Hivac$\,,
and a \emph{Weinberg angle} \txtm{\theta\in(0,\pi/2)}.
Together, these determine a `symmetry breaking' in the geometry of $\Y$,
and decompositions of the fields into several sectors.
One straightforwardly recovers the standard notions of e.w.\ theory
(the complete procedure is too intricate to be reported in this summary).
\end{mybulist}


A further subject which finds its natural fomulation
within the two-spinor setting is that of Lie derivatives of spinors and connections.
Indeed, one finds that the Lie derivatives of two-spinors and Dirac spinors,
as well as the Lie derivatives of the spinor connection and of the soldering form,
are are stricly related to one another and, together,
describe the \emph{deformations} of Einstein--Cartan--Dirac fields.\footnote{
See also the dedicated paper~\cite{Canarutto16d} for the details,
that lie ouside the scope of this presentation.} 

Two-spinor geometry also suggests a natural extension of the Higgs sector
of the electroweak theory,
in which one has a natural Lagrangian such that the self-interactions
of the extended Higgs field sum up to zero.
In turn, this suggest a possible way for explaining the standard Higgs potential
and the `breaking of dilatonic symmetry'.

\subsection{Multi-particle algebra and operator algebra}
\label{s:Multi-particle algebra and operator algebra}

Let \txtm{\Z\onto\X} be a vector bundle and denote by \txtm{\uZC^1}
the \emph{freely generated} vector space of sections \txtm{z:\X\to\Z},
that is the space of all such sections which vanish outside a finite subset of $\X$\,.
Similarly, denote by \txtm{\uZC^{\lin1}}
the freely generated vector space of sections \txtm{\zeta:\X\to\Zl}.
These spaces, which are infinite-dimensional unless the cardinality of $\X$ is finite,
can be regarded as mutually dual.\footnote{
We need not be concerned with duality in the most general acceptation.} 

The space $\uZC^1$ is a template for the space of states of one particle of some type.
The associated `$n$-particle state' space
$$\uZC^n\equiv\lozenge^n\uZC^1$$
is defined as either the symmetrized tensor product $\vee^n\uZC^1$ (\emph{bosons})
or the  antisymmetrized tensor product $\weu{n}\uZC^1$ (\emph{fermions}).
For \txtm{y\in\uZC^m,\,z\in\uZC^n}, \txtm{y\wee z\in\uZC^{m+n}} is
either $y\,{\vee}\,z$ or $y\we z$\,, as appropriate, and called
the \emph{exterior product} of $y$ and $z$\,.

Setting \txtm{\uZC^0\equiv\CC} (called the \emph{vacuum}),
the \emph{multi-particle state space}
for the particle type under consideration is defined as
$$\uZC\equiv\medoplus_{\mathclap{n=0}}^\infty\uZC^n~,$$
Namely, $\uZC$ is constituted by all formal, finite sums with arbitrarily many terms.
The similarly defined space \txtm{\uZC^\lin\equiv\medoplus_{{n=0}}^\infty\uZC^{\lin n}}
is regarded as its `dual'.
By linearity, the exterior product \txtm{(y,z)\mapsto y\wee z} can be extended
to any couple of elements in $\uZC{}$.
Hence $\uZC$ is defined similarly to the usual \emph{Fock spaces} of QFT.\footnote{
This setting suffices if one is not concerned with completions, namely with infinite sums:
our multi-particle states only contain finitely many particles,
though their number can be arbitrarily large.} 
One also has an \emph{interior product},
$$\uZC\times\uZC^\lin\to\uZC\cup\uZC^\lin:(\la,z)\mapsto\la\,|\,z~,$$
which belongs to $\uZC$ or $\uZC^\lin$ depending on which of the two factors is of higher rank.
For fermions, this is the usual interior product
\txtm{i[\la]z} of exterior algebra.
For bosons it can be defined similarly,
as tensor contraction with appropriate symmetrization and normalization
such that it fulfils the rule
$$(\zeta\wee\la)\,|\,z=\la\,|\,(\zeta\,|\,z)~,\qquad
\zeta\in\uZC^{\lin1}\,,~\la\in\uZC^\lin~.$$

A general theory of quantum particles has several particle types;
correspondingly, we consider several
multi-particle state spaces
$\uZC'$, $\uZC''$, $\uZC'''$\!, and so on.
The \emph{total state space} is now defined as
$$\uVC\equiv\uZC'\tn\uZC''\tn\uZC'''\tn\cdots=
\medoplus_{\mathclap{n=0}}^\infty\uVC^{n}~,$$
where $\uVC^{n}$, consisting of all elements of tensor rank $n$,
is the space of all states of $n$ particles of any type.
Moreover, all fermionic sectors can be described by a unique
overall antisymmetrized tensor algebra.\footnote{
If $\XC$ and $\YC$ are any two vector spaces,
then their antisymmetric tensor algebras fulfil the isomorphisms
$$\weu{p}(\XC\,{\oplus}\,\YC)\cong
\medoplus_{h=0}^p\,(\weu{p-h}\XC)\tn(\weu{h}\YC)~,\quad
(\wedge\XC)\tn(\wedge\YC)\cong\wedge(\XC\,{\oplus}\,\YC)~.$$
} 
A similar observation holds true for the bosonic sectors.
Furthermore, the mutual ordering of fermionic and bosonic sectors is regarded as inessential.

Letting the \emph{parity} (or \emph{grade}) \txtm{\grade{\phi}\in\ZZ_2}
of a monomial element \txtm{\phi\,{\in}\,\uVC}
be the number of its fermion factors $({\mathrm{mod}}\;2)$,
one gets a structure of `super-algebra' (a $\ZZ_2$-graded algebra) on $\uVC$\,.
The algebra product, denoted by $\scriptstyle\lozenge$,
is the exterior product modulo the so-called \emph{Koszul convention},
which essentially amounts to imposing anti-commutativity (or `super-commutativity').
In particular, this implies the commutativity of the multiplication
of any element by a bosonic factor.
Similarly, one constructs a `dual' space \txtm{\uVC^\lin{}\equiv\uZC^\lin{}'\tn\uZC^\lin{}''\tn\uZC^\lin{}'''\dots}
and the interior product can be extended as a map
\txtm{\uVC{}\times\uVC^\lin\to\uVC{}\cup\uVC^\lin: (\la,\psi)\mapsto\la\,|\,\psi}\,,
where the required contractions are to be performed in the appropriate tensor factors,
yielding the identities
\begin{align*}
&(\zeta\wee\xi)\,|\,\psi=\xi\,|\,(\zeta\,|\,\psi)~,
\\[6pt]
&\psi\wee\phi=(-1)^{\grade{\phi}\grade{\psi}}\phi\wee\psi
\quad\text{(anti-commutativity)},
\\[6pt]
&z\,|\,(\phi\wee\psi)=
(z\,|\,\phi)\wee\psi+(-1)^{\grade{z}\grade{\phi}}\,\phi\wee(z\,|\,\psi)~,\qquad
\phi,\psi\in\VC,~
\zeta,\xi\in\uVC^{\lin1}.
\end{align*}

The \emph{absorption} operator associated with \txtm{\zeta\in\uVC^{\lin1}}
and the \emph{emission} operator associated with \txtm{z\in\uVC^1}
are the linear maps \txtm{\uVC\to\uVC} respectively defined as
$$\abs{a}[\zeta]\phi\equiv \zeta\,|\,\phi~,\quad
\tra{a}[z]\phi\equiv z\wee\phi~,\qquad\phi\in\uVC~.$$
Similarly, one gets operators
\txtm{\abs{a}[z],\tra{a}[\zeta]:\uVC^\lin\to\uVC^\lin}
and gets \txtm{\la\,|\,\abs{a}[\zeta]\psi=(\tra{a}[\zeta]\la)\,|\,\psi}\,,
\txtm{\forall\la\in\uVC^\lin},
namely $\abs{a}[\zeta]$ and $\tra{a}[\zeta]$ are mutually transpose endomorphisms.

Let now \txtm{\abs{op}:\uVC^1\oplus\uVC^{\lin1}\to\End(\uVC)}
be the linear map characterized by
$$\abs{op}[v]\equiv\begin{cases}
\abs{a}[v]~,\quad v\in\uVC^{\lin1}~,\\
\tra{a}[v]~,\quad v\in\uVC^1~,\end{cases}$$
and consider the vector space
\txtm{\uVC^\otimes\equiv\medotimes_{n=0}^\infty(\uVC^1\oplus\uVC^{\lin1})}.
Then one gets a natural morphism
\txtm{\abs{op}:\uVC^\otimes\hookrightarrow\End(\uVC)}
of associative algebras,
defined for decomposable tensors by
$$x\tn y\tn\cdots\tn z~\longmapsto~
\abs{op}[x]\comp\abs{op}[y]\comp\cdots\comp\abs{op}[z]~,$$
and \txtm{\abs{op}[c]\equiv c\,\id} for a zero-rank tensor \txtm{c\in\CC}\,.

Let now the grades of $\abs{a}[\zeta]$ and $\tra{a}[z]$
be $\grade{\zeta}$ and $\grade{z}$, respectively,
and the grade of any composition the sum (mod~2) of the grades of all factors.
The \emph{super-bracket} (or \emph{super-commutator})
of \txtm{X,Y\in\abs{op}(\uVC^\otimes)}
is then defined by
$$\suc{X,Y}\equiv X\,Y-(-1)^{\sst\grade{X}\grade{Y}}YX$$
whenever both $X$ and $Y$ have definite grade, and extended by linearity.
In particular, for all \txtm{y,z\in\uVC^1}
and \txtm{\zeta,\xi\in\uVC^{\lin1}} one has
$$\suc{\abs{a}[\xi],\abs{a}[\zeta]}=\suc{\tra{a}[y],\tra{a}[z]}=0~,\qquad
\suc{\abs{a}[\zeta],\tra{a}[z]}=\bang{\zeta,z}\,\id~.$$


Because of the above \emph{super-commutation relations},
the morphism \txtm{\abs{op}:\uVC^\otimes\hookrightarrow\End(\uVC)}
is not a monomorphism.
On the other hand, consider the subspace
$$\uVC^\lozenge\equiv
\operatornamewithlimits{\lozenge}_{n=0}^\infty(\uVC^1\oplus\uVC^{\lin1})
\subset\uVC^\otimes~,$$
whose decomposable elements are Koszul products.
If the product's rules and the related
identifications---that is super-commutativity---%
are applied to the exchange
between elements in $\uVC$ and in $\uVC^\lin$ as well,
then one gets  the identification
$$\uVC^\lozenge\cong\uVC\tn\uVC^\lin~,$$
where, in each decomposable element, all `covariant' factors
are on the right of any `contravariant' factors.
This is called \emph{normal ordering}.
The image
$$\uOC\equiv\abs{op}(\uVC^\lozenge)\subset\End(\uVC)~,$$
is our fundamental operator space.
A bilinear product \txtm{\uOC\times\uOC\to\uOC}
can be defined as composition together with super-commutative
normal reordering in each decomposable term.
This renders $\uOC$ a super-commutative $\ZZ_2$-graded algebra.

\subsection{Quantum bundles and quantum fields}
\label{s:Quantum bundles and quantum fields}

In my opinion, that part of elementary particle theory
which is unquestionably rooted in actual physics
can be actually introduced without making essential use of quantum fields.
I tend to regard the notion of quantum field as auxiliary rather than fundamental,
entangled with particle physics for historical reasons;
and I suspect that the issue of finding a full-fledged covariant
approach to quantum fields in curved spacetime might eventually turn out to be pointless.
Nevertheless, since they are needed in order to cope with the literature,
I studied a precise mathematical approach to quantum bundles
and quantum fields, including ghosts, BRST symmetry,
and the so-called `anti-field' formalism.
The geometry of quantum bundles and their jet prolongations can be developed in
terms of F-smoothnes.


As an intermediate step, we need the notion of a distributional bundle, that is,
a bundle over $\M$ whose fibres are 
distributional spaces.
In general, the finite-dimensional geometric structure
underlying functional bundles~\cite{KolarModugno98,Canarutto00a,Canarutto01,
CabrasJanyskaKolar04,CabrasJanyskaKolar06,Canarutto15a}
is that of a two-fibered bundle \txtm{\Z\onto\Y\onto\X}.
If \txtm{\Z\onto\Y} is a vector bundle, then for any \txtm{x\in\X}
one obtains the vector space \txtm{\DC_x(\Y,\Z)} of all section-distributions~\cite{SchwartzL66}
\txtm{\Y\!\!_x\to\Z_x}\,.
A smooth bundle structure on
the fibred set \txtm{\DC(\Y,\Z)\equiv\bigsqcup_{x\in\X}\DC_x(\Y,\Z)\onto\X}
can be assigned, exploiting Fr\"olicher's notion of smoothness,\footnote{
If $\XC$ is any set, then a family $\CCal$ of curves \txtm{\RR\to\XC}
determines the family $\FC\CCal$ of maps \txtm{\XC\to\RR}
fulfilling \txtm{f\in\FC\CCal} if and only if \txtm{f\comp c:\RR\to\RR} is smooth.
Conversely, a set $\FC$ of functions \txtm{\XC\to\RR} determines
a set $\CCal\FC$ of curves in $\XC$ by the same requirement.
Any set of curves, or any set of functions,
generates such an F-smooth structure on $\XC$.
This notion of smoothness, which was introduced by Fr\"olicher~\cite{Froelicher82},
is compatible with the standard one in finite-dimensional manifolds;
moreover, it behaves naturally with regard to inclusions and Cartesian products,
so it yields a convenient general setting for dealing
with functional spaces and functional bundles~\cite{FroelicherKriegl88,JanyskaModugno20,
KrieglMichor97,KolarModugno98,CabrasJanyskaKolar04,CabrasJanyskaKolar06}.} 
by selecting, as the set $\CCal\!_{\sst\DC}$ of all F-smooth curves,
the set of all local maps \txtm{c:\RR\to\DC} such that the map
\txtm{\bang{c,u}:\RR\to\CC:t\mapsto\bang{c(t),u}}
is smooth for any test element \txtm{u}
(a smooth section \txtm{\Y\to\Z} with compact support).

By replacing $Z$ with \txtm{\weu{n}\VO^*\Y\ten{\Y}\Z}\,,
where $n$ is the fibre dimension of \txtm{\Y\onto\X}\,,
one gets an F-smooth bundle of generalized \emph{densities}.
If the fibres of \txtm{\Y\onto\X} are smoothly orientable then one may choose
a positive sub-bundle \txtm{\VV\equiv(\weu{n}\VO\Y)^+},
and gets the F-smooth bundle \txtm{\DC(\Y,\VV^{-1/2}\tn\Z)\onto\X}
of \emph{$\Z$-valued semi-densities}.
This is especially convenient as a template for bundles of quantum states,
since when a fibre Hemitian structure of \txtm{\Z\onto\Y} is assigned
one gets a bundle of \emph{rigged Hilbert spaces} \txtm{\DCo\subset\HC\subset\DC},
where the fibres of \txtm{\DCo\onto\X} and \txtm{\HC\onto\X} are generated, respectively,
by test semi-densities and square-integrable semidensities.

We are mainly interested in the case when \txtm{\Y\equiv\Pm\onto\M},
the \emph{future mass-shell bundle} for \txtm{m\in\LL\!^{-1}},
that is the sub-bundle of \txtm{\TS\M\onto\M}
whose fibres consist of all future-oriented covectors of Lorentzian pseudo-norm $m$\,.

Let us denote the ensuing F-smooth bundle of $\Z$-valued semi-densities by \txtm{\ZC^1}.
Then, the vector bundles
\[ \ZC^n\equiv\lozenge^n\ZC^1\onto\M~,\qquad
\ZC\equiv\medoplus_{\mathclap{n=0}}^\infty\ZC^n\onto\M~,\]
are constructed, fibrewise, with the same procedure
as in~\Sec\ref{s:Multi-particle algebra and operator algebra}.
Furthermore, one has the sub-bundles \txtm{\uZC^n\subset\ZC^n}, \txtm{\uZC\subset\ZC},
whose fibres consist of finite linear combinations
of \emph{Dirac-type} semi-densities.
The latter are expressable as \txtm{\upeta^{-1/2}\tn\updelta[p]\tn z}\,, where
\begin{mybulist}
\item
The density \txtm{\upeta:\M\to\DC(\Pm\,,\!\VV^{-1})} is the metric-induced volume form
on the fibres of \txtm{\Pm\!\onto\!\M}
\item
\txtm{p:\M\to\Pm} is a smooth section
\item
\txtm{\updelta[p]:\M\to\DC(\Pm\,,\!\VV^{-1})} is, at each \txtm{x\in\M},
the Dirac density \txtm{\updelta[p(x)]} on $(\Pm)_x$
\item
\txtm{z:\Pm\to\Z} is a smooth section (only its values along the image of $p$ matter).
\end{mybulist}

The bundle \txtm{\uZC\onto\M} is relevant under two respects.
First, the inclusion \txtm{\uZC\,{\subset}\,\ZC} is \emph{dense}~\cite{SchwartzL66};
second, the corrispondence \txtm{\upeta^{-1/2}\tn\updelta[p]\tn z\leftrightarrow z(p)}
determines a fibred isomorphism between $\uZC$
and the bundle whose fibres are freely generated spaces of sections \txtm{\Pm\to\Z}
(\Sec\ref{s:Multi-particle algebra and operator algebra}).
Namely, multi-particle algebra and graded operator algebra can be readily adapted
to distributional spaces and bundles.
Considering several sectors,
by constructions similar to~\Sec\ref{s:Multi-particle algebra and operator algebra}
we  get the bundle \txtm{\VC^\lozenge\cong\VC\tn\VC^\lin\onto\M},
with \txtm{\VC\equiv\ZC'\tn\ZC''\tn\ZC'''\tn\cdots}
as well as its `underlined' counterparts (finitely generated by semi-densities of Dirac type).
Furthermore, we get the operator-algebra bundle \txtm{\OC\equiv\abs{op}(\VC^\lozenge)\onto\M};
its elements, in general, are linear morphisms \txtm{\VCo\to\VC},
where \txtm{\VCo\subset\VC} is the sub-bundle generated by test semi-densities
(Dirac deltas cannot be contracted with arbitrary distributions),
but may admit extensions.

Consider the simpler case in which the `internal' bundle
is actually a vector bundle \txtm{\E\onto\M}, so that \txtm{Z=\Pm\cart\M\E}.
The associated \emph{quantum bundle} is then defined as the vector bundle
\[ \EC\equiv\OC\ten{\M}\E\onto\M~,\]
where \txtm{\OC\onto\M} is the operator-algebra bundle previously described
(it is not difficult to extend this `bundle quantization' to the general situation
of two-fibred internal bundles, which notably includes the electron bundle
\txtm{\W^+\onto\Pm} and the positron bundle \txtm{\Wc{}^-\onto\Pm}).
A Fr\"olicher-smooth structure of \txtm{\EC\onto\M} can be readily introduced,
and all the basic differential geometric notions for classical bundles
can be naturally extended to quantum bundles.
A few observations are in order.

\begin{mybulist}
\item
A frame of \txtm{\E\onto\M} can be regarded as a frame of \txtm{\EC\onto\M}:
the components of fibre elements are elements in $\OC$,
which can be regarded as `quantum numbers'.
A polynomial \txtm{\E\to\CC} in the fibre coordinates
determines a `quantum polynomial' \txtm{\EC\to\OC};
more general quantum functions are not actually needed in gauge field theory.
Partial derivatives with respect to fibre coordinates are then well-defined
algebraic operations.
\item
In order to recover the standard formalism one has to assume
that \txtm{\OC\onto\M} is a trivial bundle.
A special trivialization is determined by the choice of an observer
(a frame may be either associated with the observer or not).
Then $\OC$ can be viewed as a fixed $\ZZ_2$-graded algebra,
and a linear connection of \txtm{\E\onto\M} yields a linear connection of \txtm{\EC\onto\M}.
\item
In a local frame, the components of a \emph{quantum field} \txtm{\phi:\M\to\EC}
are $\OC$-valued.
A Lagrangian density \txtm{\JO\EC\to\OC\tn\weu4\TS\M}
can be written as \txtm{\Lambda=\uplambda\,\upeta}\,,
where $\upeta$ is the metric volume form of spacetime
and $\uplambda$ is a quantum polynomial in the components
of the field and its covariant derivatives
(the gauge field itself must be quantized, so that a gauge is needed).
The field equations are derived as in the classical theory,
but attention to signs must be payed since $\OC$ is $\ZZ_2$-graded.
\item
The existence of sections \txtm{\phi:\M\to\EC} fulfilling
the complete field equations, with interactions, is far from guaranteed,
but this is not actually an issue in perturbative particle physics.
\item
Given an observer and a frame adapted to it one may recover the usual notions
of absorption and emission operators, free quantum fields, and so on.
\end{mybulist}

In a theory with several matter sectors,
each one is assumed to be either \emph{bosonic} or \emph{fermionic}.
This means that $\EC$ is not obtained by tensorializing $\E$
by the whole $\ZZ_2$-graded operator algebra, \txtm{\OC=\OC_\gradezero\oplus\OC_\gradeone}\,,
but rather as either \txtm{\OC_\gradezero\tn\E} or \txtm{\OC_\gradeone\tn\E}, respectively.
Considering the exchanged parity in each sector gives rise to the notion of \emph{anti-field}
and to the Batalin--Vilkovisky algebra.

Furthermore, in the quantum theory derived from a classical field theory,
one also has to include additional \emph{ghost} fields.
For example, consider the theory derived from a classical theory
of a fermion field\footnote{
Hre, the fermion field may have further internal structure besides spin.} 
\txtm{\psi:\M\to\W\ten\M\E}
and a bosonic gauge field \txtm{{\st A}:\M\to\TS\M\ten\M\Lie}\,,
where \txtm{\Lie\subset\End\E}.
Then, in addition to the independent adjoint fermion field
\txtm{\bar\psi:\M\to\Wl\ten\M\E^\lin},
one also has the fermionic \emph{ghost} and \emph{anti-ghost} fields,
\txtm{\gh:\M\to\Lie} and \txtm{\agh:\M\to\Lie^*},
and the bosonic \emph{Nakanishi--Lautrup field} \txtm{{\st N}:\M\to\Lie}\,.

In general, a vertical symmetry of the Lagrangian is a fibred morphism
\txtm{v:\JO\EC\to\VO\EC} (a \emph{generalized vector field})
fulfilling\footnote{
The operator $\delta[v]$ acts on horizontal forms
\txtm{\alpha:\JO_k\EC\to\OC\tn\weu{q}\TS\M} as
\txtm{\delta[v]\alpha\equiv\check\dO\alpha\pint v_{\sst(k)}:\JO_{k+1}\EC\to\OC\tn\weu{q}\TS\M},
where \txtm{\check\dO\alpha} is the fibre differential of $\alpha$
and the generalized vector field \txtm{v_{\sst(k)}:\JO_{k+1}\EC\to\VO\JO_k\EC}
is the holonomic restriction of \txtm{\JO_kv:\JO_k\JO\EC\to\JO_k\VO\EC},
taking the natural isomorphism \txtm{\JO_k\!\VO\EC\cong\VO\JO_k\EC} into account.} 
\txtm{\delta[v]\Lambda=\dH\nu}\,,
where \txtm{\nu:\JO\EC\to\OC\tn\weu3\TS\M} and $\dH$ denotes the horizontal differential.
It turns out that the Lagrangian of the above sketched theory,
besides the symmetries of the classical theory,
has a further symmetry, which is a special case of the \emph{BRST symmetry}.\footnote{
The details of the expressions of the Lagrangian and of the symmetry are not included here,
as they are somewhat involved.} 
In this case the components of $v$, in each sector, have the sector's parity.
The \emph{BRST transformation},
acting on horizontal forms \txtm{\alpha:\JO_k\EC\to\OC\tn\weu{q}\TS\M},
is defined in term of an arbitrarily chosen element \txtm{\theta\in\OC_{\grade1}} by
\[ \theta\,\brstS\alpha\equiv\delta[v]\alpha~,\qquad
\delta[v]\alpha:\JO_{k+1}\EC\to\OC\tn\weu{q}\TS\M~,\]
and turns out to be nilpotent (\txtm{\brstS^2=0}).


\subsection{Quantum particles}
\label{s:Quantum particles}

In recent work, and especially in the aforementioned book~\cite{Canarutto2020},
I proposed an approach to the physics of quantum particles
that essentially dispenses with quantum fields.
Indeed, although the descriptions of quantum particles and quantum fields both stem
from the underlying classical bundle geometry, quantum fields can be actually sidestepped.

First, a chosen time function is needed.
This requirement is common to all approaches, though it is seen as a drawback---%
as it `breaks Lorentz invariance'.
In geometric terms, it can be implemented in various ways.
My proposal starts from the notion of a \emph{detector},
that is a one-dimensional timelike submanifold \txtm{\T\subset\M}.
Along $\T$, one gets the orthogonal splitting
\[ (\TM)_{\sst\T}=\TO\T\dir\T(\TM)_{\sst\T}^\sbot \]
into timelike and spacelike sub-bundles,
and a similar splitting of the cotangent bundle.
Particle spins and momenta can be Fermi-transported along $\T$,
while for other internal degrees of freedom, not soldered to spacetime,
one needs a gauge.\footnote{
A background connection, also needed in field quantization.} 
Accordingly, one constructs \emph{free-particle states};
these, together with particle interactions,
constitute the building blocks of the dynamics of many-particle quantum systems,
yielding a `perturbative' formalism in momentum space which can be viewed
as a sort of complicated `clock' carried by the detector.
By choosing suitable classical frames,
and considering the associated quantum \emph{generalized frames}
(which are Dirac-type semi-densities),
one is able to recover the standard scattering matrix computations
in terms of free states and the interactions among them.\footnote{
The examples offered in the book~\cite{Canarutto2020} are meant to support
the philosophy of my approach,
with no claim of constituting a comprehensive treatment of this subject.
The details do not fit into this brief summary.} 

Similarly, the quantum interaction is constructed from the classical interaction---which
is in turn a natural byproduct of the underlying classical geometry---coupled with
a `quantum ingredient' which is a special generalized semi-density on the space
of particle momenta.\footnote{
Essentially, the Dirac delta of the sum of momenta in the usual formulation.} 

In a curved spacetime background,
the precise correspondence between momentum and position representations is lost,
and I tend to view the former as more fundamental.
The latter can be recovered \emph{locally},
using the observation that exponentiation yields a family \txtm{\bigl\{\X_t\,,t\in\T\bigr\}}
of spacelike submanifolds orthogonal to $\T$;
hence, one has a time\txtm{\,{\oplus}\,}space splitting in a neighbourhood of $\T$.

I also want to point out to an intriguing consequence of the two-spinor formalism,
associated with the idea that gauge fields and gauge particles
are to be treated as distinct from connections
(though the two notions are strictly related).
By describing the interactions between fermions and photons in two-spinor terms,
and using the free-particle relations between momentum and spin,
gauge freedom can be recovered in a purely algebraic way.
I view this result as supporting the idea that the relation between
classical geometry and the quantum description could be inverted:
the former could be obtained from the latter.
Namely, \emph{the system defines the geometry}~\cite{Penrose71}
and reality is fundamentally discrete,
while notions related to continuity should be recovered as conveniences
in the description of sufficiently complex systems.
Ideas of this kind have been around for some time
and have inspired attempts at serious theories~%
\cite{Penrose71,Verlinde2010}.

My long-term goal\footnote{
Maybe for my next reincarnation} 
is somewhat radical, since I would like to achieve
a fully \emph{discrete}, \emph{relational} theory of spacetime and matter.
I speculate~\cite{Canarutto11a} that physical reality \emph{is} fundamentally a network,
whose nodes and edges may be called \emph{events} and \emph{particles}, respectively.
Approximate geometric relations among edges will emerge
in a sufficiently large portion of the network;
information about some edges will constrain the network's state,
allowing a probabilistic description of the missing information.
In this scenario of emerging geometry
we should regard smooth manifolds and bundles,
as well as connections and algebraic structures
(including the spacetime metric),
as secondary macroscopic notions.
Hence spacetime, gravity (\emph{not} quantum gravity),
and quantum mechanics could all emerge from a more fundamental, discrete theory.

If such a program could be fulfilled, then classical notions
would take the role of `mean field' background properties of physical systems.
In particular, this would be true for connections of classical bundles,
implying that the relation between gauge particles and connections
should be thought over
and need not be a precise correspondence.
We may then expect gauge freedom
expressed in two-spinor terms to play a relevant role.
Indeed, it would be a very interesting outcome if
\emph{any} spacetime-related notion turned out to be founded on two-spinors.

Finally, let me observe how a possible easing of tensions
among notions relevant to quantum physics
could be achieved by acknowledging their diverse roles and levels of importance.
We should seek no mandatory unification of everything on the same footing;
above all, gravitation is \emph{not} assimilable to the other interactions.
Moreover, the complementarity of momentum
and spatial position representations
is of partial and limited scope:
the former is more directly linked to internal particle structure
and quantum interactions,
while the latter is related to emerging geometry and the notion
of a quantum field---which can be described as a section of a quantum bundle
and not regarded as truly fundamental.



\begin{thebibliography}{10}
%
\bibitem{Canarutto2020}
D.\ Canarutto:
\emph{Gauge field theory in natural geometric language},
Oxford University Press, Oxford (2020).
%
\bibitem{ChoquetBruhatDeWittMorette}
Y.\ Choquet-Bruhat and C.\ DeWitt-Morette:
\emph{Analysis, manifolds and physics},
North-Holland, Amsterdam (1982).
%
\bibitem{Penrose2016}
 R.\ Penrose:
\emph{Fashion faith and fantasy in the New Physics of the Universe},
Princeton University Press, Princeton and Oxford (2016).
%
\bibitem{Canarutto98}
D.\ Canarutto:
`Possibly degenerate tetrad gravity and Maxwell-Dirac fields',
\emph{J.\ Math.\ Phys.}\ {\bf 39}, 4814--23 (1998).
%
\bibitem{Canarutto00a}
D.\ Canarutto:
`Smooth bundles of generalized half-densities',
\emph{Arch.\ Math.\ Un.\ Brunensis} {\bf 36}, 111--24 (2000).
%
\bibitem{Canarutto00b}
D.\ Canarutto:
`Two-spinors, field theories and geometric optics in curved spacetime',
\emph{Acta Appl.\ Math.}\ {\bf 62},  187--224 (2000).
%
\bibitem{Canarutto01}
D.\ Canarutto:
`Generalized densities and distributional adjoints of natural operators',
\emph{Rend.\ Semin.\ Mat.\ Univ.\ Pol.\ Torino} {\bf 59},  27--36 (2001).
%
\bibitem{Canarutto04a}
D.\ Canarutto:
`Connections on distributional bundles',
\emph{Rend.\ Semin.\ Mat.\ Univ.\ Padova} {\bf 111}, 71--97 (2004).
%
\bibitem{Canarutto05}
D.\ Canarutto: 
`Quantum bundles and quantum interactions',
\emph{Int.\ J.\ Geom.\ Met.\ Mod.\ Phys.}\ {\bf 2}, 895--917 (2005).
%
\bibitem{Canarutto07}
D.\ Canarutto: 
`{}``Minimal geometric data'' approach to
Dirac algebra, spinor groups and field theories',
\emph{Int.\ J.\ Geom.\ Met.\ Mod.\ Phys.}\ {\bf 4},  1005--40 (2007).
%
\bibitem{Canarutto09}
D.\ Canarutto:
`Fermi transport of spinors and free QED states in curved spacetime',
\emph{Int.\ J.\ Geom.\ Met.\ Mod.\ Phys.}\ {\bf 6}, 805--24 (2009).
%
\bibitem{Canarutto11}
D.\ Canarutto:
`Tetrad gravity, electroweak geometry and conformal symmetry',
\emph{Int.\ J.\ Geom.\ Met.\ Mod.\ Phys.}\ {\bf 8}, 797--819 (2011).
%
\bibitem{Canarutto12a}
D.\ Canarutto: 
`Positive spaces, generalized semi-densities and quantum interactions',
\emph{J.\ Math.\ Phys}.\ {\bf53}, 032302 (2012).
%
\bibitem{Canarutto15a}
D.\ Canarutto: 
`Fr\"olicher-smooth geometries, quantum jet bundles and BRST symmetry',
\emph{J.\ Geom.\ Phys.}\ {\bf88}, 113--28 (2015).
%
\bibitem{Canarutto15b}
D.\ Canarutto:
`Natural extensions of electroweak geometry and Higgs interactions',
\emph{Ann.\ H.\ Poincar\'e} {\bf16},  2695--711 (2015).
%
\bibitem{Canarutto16a}
D.\ Canarutto:
`Special generalized densities and propagators: a geometric account',
\emph{Int.\ J.\ Geom.\ Met.\ Mod.\ Phys.}\ {\bf13}, 1530004 (2016).
%
\bibitem{Canarutto16b}
D.\ Canarutto:
`On the geometry of ghosts',
\emph{Rep.\ Math.\ Phys.}\ {\bf78}, 123--56 (2016).
%
\bibitem{Canarutto16c}
D.\ Canarutto:
`Overconnections and the energy-tensors of gauge and gravitational fields',
\emph{J.\ Geom.\ Phys.}\ {\bf106}, 192--204 (2016).
%
\bibitem{Canarutto16d}
D.\ Canarutto: 
`Two-spinor tetrad and Lie derivatives of Einstein-Cartan-Dirac fields',
\emph{Arch.\ Math.\ Un.\ Brunensis} {\bf54}, 205--26 (2018).
%
\bibitem{Canarutto16e}
D.\ Canarutto:
`Covariant-differential formulation of Lagrangian field theory',
\emph{Int.\ J.\ Geom.\ Met.\ Mod.\ Phys.}\ {\bf15}, 1530004 (2018).
%
\bibitem{Canarutto16f}
D.\ Canarutto: 
`A first-order Lagrangian theory of fields with arbitrary spin',
\emph{Int.\ J.\ Geom.\ Met.\ Mod.\ Phys.}\ {\bf15}, 1850088 (2018).
%
\bibitem{Canarutto18a}
D.\ Canarutto: 
`On the notions of energy tensors in tetrad-affine gravity',
\emph{Grav.\ Cosmol.}\ {\bf24}, 122--8 (2018).
%
\bibitem{CanaruttoJadczyk97a}
D.\ Canarutto and A.\ Jadczyk:
`Fundamental geometric structures for the Dirac equation in General Relativity',
\emph{Acta Appl.\ Math.}\ {\bf 50}, 59--92 (1998).
%
\bibitem{CanaruttoMinguzzi2019}
D.\ Canarutto and E.\ Minguzzi:
`The distance formula in algebraic spacetime theories',
\emph{J.\ Phys.\ Conf.\ Ser.}\ {\bf 1275}, 012045 (2019).
%
\bibitem{PenroseRindler84}
R.\ Penrose and W.\ Rindler:
\emph{Spinors and space-time.
{I}: two-spinor calculus and relativistic fields},
Cambridge University Press, Cambridge (1984).
%
\bibitem{PenroseRindler88}
R.\ Penrose and W.\ Rindler:
\emph{Spinors and space-time.
{II}: spinor and twistor methods in space-time geometry},
Cambridge University Press, Cambridge (1988).
%
\bibitem{JanyskaModugnoVitolo10}
J.\ Jany{\v s}ka, M.\ Modugno, R.\ Vitolo:
`An algebraic approach to physical scales',
\emph{Acta Appl.\ Math.}\ {\bf 110}, 1249--76 (2010).
%
\bibitem{CabrasJanyskaKolar04}
A.\ Cabras, J.\ Jany{\v s}ka,  I.\ Kol\'a\v{r}:
`Functorial prolongations of some functional bundles',
\emph{Annales Acad.\ Paed.\ Cracoviensis, Stud.\ Math.\ IV}
{\bf 23}, 17--30 (2004).
%
\bibitem{CabrasJanyskaKolar06}
A.\ Cabras, J.\ Jany{\v s}ka,  I.\ Kol\'a\v{r}:
`On the geometry of the variational calculus on some functional bundles',
\emph{Note di Mat.}\ {\bf 26}, 51--66 (2006).
%
\bibitem{KolarModugno98}
I.\ Kol\'a\v{r} and M.\ Modugno:
`The Fr\"olicher--Nijenhuis bracket on some functional spaces',
\emph{Ann.\ Pol.\ Math.}\ {\bf 68}, 97--106 (1998).
%
\bibitem{SchwartzL66}
L.\ Schwartz:
\emph{Th\'eorie des distributions},
Hermann, Paris (1966).
%
\bibitem{Froelicher82}
 A.\ Fr\"olicher:
\emph{Smooth structures},
LNM {\bf962}, Springer-Verlag, 69--81 (1982).
%
\bibitem{FroelicherKriegl88}
 A.\ Fr\"olicher and A.\ Kriegl:
\emph{Linear spaces and differentiation theory},
John Wiley~\&~Sons, New York (1988).
%
\bibitem{JanyskaModugno20}
J.\ Jany{\v s}ka and M.\ Modugno:
`Smooth and F–smooth systems',
arXiv:2002.11983 [math.DG].
%
\bibitem{KrieglMichor97}
A.\ Kriegl and P.\ Michor:
\emph{The convenient setting of global analysis},
American Mathematical Society (1997).
%
\bibitem{Penrose71}
R.\ Penrose:
`Angular momentum: an approach to combinatorial space-time',
in \emph{Quantum Theory and Beyond---%
essays and discussions arising from a colloquium},
T.\ Bastin editor,
Cambridge University Press, Cambridge, 151--80 (1971).
%
%
\bibitem{Verlinde2010}
E.P.\ Verlinde:
`On the origin of gravity and the laws of Newton', arXiv:1001.0785v1 (2010).
%
%
\bibitem{Canarutto11a}
D.\ Canarutto: `Nature's software',
essay presented for the 2011 contest,
`Is Reality Digital or Analog?',
of the Foundational Questions Institute (FQXi),\\
http://fqxi.org/community/forum/topic/831.
%
\end{thebibliography}
\end{document}